# Orbital and spin components of the angular momentum of a general electromagnetic field

## A. M. Stewart

*Department of Theoretical Physics, The Research School of Physical Sciences and Engineering, The Australian National University, Canberra, ACT 0200, Australia.*

## 1. Introduction

The notion of light containing orbital and spin components of angular momentum has become of interest in recent years and a direct decomposition into the two components has been found for paraxial rays [1-4]. However, a simple, plausible and general derivation of the components from the expression for the angular momentum $J(t)$ of the classical electromagnetic field in terms of the electric $E(x,t)$ and magnetic field $B(x,t)$

$$J(t) = \frac{1}{4\pi c} \int d^3x \, \mathbf{x} \times [E(x,t) \times B(x,t)] \qquad (1)$$

(Gaussian units, bold font denotes a three-vector) seems to be lacking. Some authors [5, 6], used decompositions that were not manifestly gauge invariant. Other decompositions lack generality [2, 3, 7]. In this paper, by applying the vector decomposition theorem of Helmholtz to the electric field, we obtain a decomposition of the angular momentum of the classical electromagnetic field into orbital and spin components that is explicit, quite general and manifestly gauge invariant throughout because it involves the fields only and not the potentials.

## 2. Helmholtz decompositions

The vector decomposition theorem of Helmholtz [8-11] states that any 3-vector field $E(x)$ that vanishes at spatial infinity can be expressed uniquely as the sum of two terms

$$E(x,t) = -\nabla_x f(x,t) + \nabla_x \times F(x,t) \qquad , \qquad (2)$$

where $\nabla_x$ is the gradient operator with respect to $x$ and the scalar $f$ and vector $F$ potential functions are

$$f(x,t) = \int d^3y \, \frac{\nabla_y \cdot E(y,t)}{4\pi |x-y|} \quad \text{and} \quad F(x,t) = \int d^3y \, \frac{\nabla_y \times E(y,t)}{4\pi |x-y|} \qquad . \qquad (3)$$

The first term of (2) is called the longitudinal part, the second the transverse part. Some authors have questioned if the Helmholtz theorem applies to fields that vary with time and it has been confirmed elsewhere [10-12] that it does. If the field is the electromagnetic $E$ field then, after using Maxwell equations, the two potential functions in (2) become

$$f(\mathbf{x},t) = \int d^3y \frac{\rho(\mathbf{y},t)}{|\mathbf{x}-\mathbf{y}|} \quad \text{and} \quad \mathbf{F}(\mathbf{x},t) = -\frac{\partial}{\partial t}\int d^3y \frac{\mathbf{B}(\mathbf{y},t)}{4\pi c|\mathbf{x}-\mathbf{y}|} \tag{4}$$

where $\rho(\mathbf{x},t)$ is the electric charge density and $\mathbf{B}$ the magnetic Maxwell field. We see that the Helmholtz theorem decomposes the Maxwell electric field into gradients of the electromagnetic Coulomb gauge potentials $\mathbf{E} = -\nabla f - \partial \mathbf{A}_t/\partial ct$. The expression for Coulomb gauge electromagnetic vector potential $A_t$,

$$\mathbf{A}_t(\mathbf{x},t) = \nabla_x \times \int d^3y \frac{\mathbf{B}(\mathbf{y},t)}{4\pi |\mathbf{x}-\mathbf{y}|} \tag{5}$$

was obtained previously [10, 11] by making a Helmholtz decomposition of the general electromagnetic vector potential. The term $\mathbf{A}_t$ is the irreducible part of the vector potential that encodes all the information about the magnetic field [11]. The pure gauge term of the vector potential, which is the gradient of a scalar field, does not encode any physical information. When the electromagnetic $\mathbf{E}$ field is decomposed in this way one part of it, that involving the gradient in equation (2), is necessarily associated with the presence of electric charge (bound or b) and one part, that involving the curl in equation (2) (free or f), is not necessarily associated with the presence of electric charge. In this sense, the fields in an irregularly shaped metal cavity will be said to be *free* although they do not have the form of plane waves. The categorisation applies to any physical quantities that depend on $\mathbf{E}$, such as the angular momentum.

As well, the electromagnetic $\mathbf{B}$ field may be decomposed by means of the Helmholtz theorem. Because $\nabla \cdot \mathbf{B} = 0$, there is only one term in the decomposition and, with the use of the inhomogeneous Maxwell equation

$$\nabla \times \mathbf{B} = \frac{4\pi}{c}\mathbf{j} + \frac{1}{c}\frac{\partial \mathbf{E}}{\partial t}, \tag{6}$$

where $\mathbf{j}$ is the electric current density, we get

$$\mathbf{B}(\mathbf{x},t) = -\frac{1}{c}\int d^3y\, \mathbf{j}(\mathbf{y},t) \times \nabla_x \frac{1}{|\mathbf{x}-\mathbf{y}|} - \frac{1}{4\pi c}\int d^3y \frac{\partial \mathbf{E}(\mathbf{y},t)}{\partial t} \times \nabla_x \frac{1}{|\mathbf{x}-\mathbf{y}|}. \tag{7}$$

The first term is an instantaneous Biot-Savart term, the second term accounts for time dependence of the fields. These Helmholtz decompositions all have the feature that they are formally instantaneous in time.

### 3. Angular momentum of the classical electromagnetic field

To calculate the angular momentum of the classical electromagnetic field we decompose the electric field according to (2) and (4) and put the result into (1).

The contribution of the free field is given by the term in (2) that contains the vector potential field $\mathbf{F}$. With the use of a standard vector identity we expand the vector product as

$$(\nabla_x \times F) \times B = (B \cdot \nabla_x)F - \sum_{r=1}^{3} B^r \nabla_x F^r \quad . \tag{8}$$

The first term contributes to the angular momentum an amount

$$J_{fs} = \frac{1}{4\pi c} \int d^3x \, x \times (B \cdot \nabla_x) F \quad , \tag{9}$$

or, in components,

$$J_{fs}^i = \frac{1}{4\pi c} \sum_{r,j,k=1}^{3} \varepsilon^{ijk} \int d^3x \, x^j B^r \frac{\partial}{\partial x^r} F^k \quad , \tag{10}$$

where $\varepsilon^{ijk}$ is the Levi-Civita tensor of rank 3. We do a partial integration with respect to $x^r$, assuming that boundary terms vanish, to get

$$J_{fs}^i = -\frac{1}{4\pi c} \sum_{r,j,k=1}^{3} \varepsilon^{ijk} \int d^3x \, F^k (\delta_{jr} B^r + x^j \frac{\partial}{\partial x^r} B^r) \quad , \tag{11}$$

where $\delta_{jr}$ is the Kronecker delta. The second term of (11) vanishes from $\nabla \cdot B = 0$ and the first term gives

$$J_{fs}^i = -\frac{1}{4\pi c} \sum_{j,k=1}^{3} \varepsilon^{ijk} \int d^3x \, B^j F^k \quad , \tag{12}$$

which is the spin component of the angular momentum, in vector form

$$J_{fs} = \frac{1}{4\pi c} \int d^3x \, F \times B \tag{13}$$

or explicitly

$$J_{fs} = \frac{1}{(4\pi c)^2} \int d^3x \int d^3y \, \frac{B(x,t)}{|x-y|} \times \frac{\partial B(y,t)}{\partial t} \quad . \tag{14}$$

The second term of (8) cannot, by repeated partial integration, be cast into a form that does not depend linearly on the vector $x$. Accordingly, it gives the orbital component of the angular momentum of the free field

$$J_{fo} = -\frac{1}{4\pi c} \int d^3x \, x \times \sum_{r=1}^{3} B^r \nabla_x F^r \quad . \tag{15}$$

By substituting for $F$ from (4) and explicitly taking the gradient this may be expressed as

$$J_{fo} = \frac{1}{(4\pi c)^2} \int d^3x \int d^3y \, [B(x,t) \cdot \frac{\partial B(y,t)}{\partial t}] \frac{x \times y}{|x-y|^3} \quad . \quad (16)$$

Equation (14) and (16) demonstrate the decomposition of the angular momentum into the two components.

## 4. Conclusion

We have obtained a decomposition of the angular momentum of the classical electromagnetic field into orbital and spin components that is general and manifestly gauge invariant. This is done by decomposing the electric field into its longitudinal and transverse parts by means of the Helmholtz theorem. The orbital and spin components of the angular momentum of any specified electromagnetic field can be found from this prescription. Applications to specific field configurations are made elsewhere [13-17].